\documentclass{optica-article}

\journal{opticajournal} 

\articletype{Research Article}

\usepackage{lineno}

\begin{document}

\title{Reflection-mode diffraction tomography of multiple-scattering samples on a reflective substrate from intensity images}

\author{Tongyu Li,\authormark{1} Jiabei Zhu,\authormark{1} Yi Shen,\authormark{1} and Lei Tian\authormark{1,2,*}}

\address{\authormark{1}Department of Electrical and Computer Engineering, Boston University, Boston, Massachusetts 02215, USA\\
\authormark{2}Department of Biomedical Engineering, Boston University, Boston, Massachusetts 02215, USA}

\email{\authormark{*}leitian@bu.edu} 


\begin{abstract*}
We introduce a novel reflection-mode diffraction tomography technique that enables simultaneous recovery of forward and backward scattering information for high-resolution 3D refractive index reconstruction. 
Our technique works by imaging a sample on a highly reflective substrate and employing a novel multiple-scattering model and reconstruction algorithm.
It combines the modified Born series as the forward model, Bloch and perfect electric conductor boundary conditions to handle oblique incidence and substrate reflections, and the adjoint method for efficient gradient computation in solving the inverse-scattering problem.
We validate the technique through simulations and experiments, achieving accurate reconstructions in samples with high refractive index contrasts and complex geometries. Forward scattering captures smooth axial features, while backward scattering reveals complementary interfacial details. Experimental results on dual-layer resolution targets, 3D randomly distributed beads, phase structures obscured by highly scattering fibers, fixed breast cancer cells, and fixed \emph{C. elegans} demonstrate its robustness and versatility. This technique holds promise for applications in semiconductor metrology and biomedical imaging.
\end{abstract*}

\section{Introduction}

Diffraction tomography (DT) is a well-established, label-free, and non-destructive imaging technique that enables the quantitative characterization of 3D refractive index (RI) distributions in samples. It has been widely applied in biomedical imaging for monitoring cellular dynamics~\cite{kim2014white} and analyzing tissue morphology~\cite{merola2017tomographic}, and it is now increasingly explored for applications in metrology and inspection~\cite{kim2016large, kang2023accelerated, aidukas2024high}. Despite its versatility, traditional DT relies primarily on transmission-mode measurements, where backscattering is considered negligible, and only forward-scattered signals are used for reconstruction. This approach faces fundamental limitations, including limited axial resolution due to the missing cone problem and poor sensitivity to lateral interfaces due to the absence of high axial spatial frequency information in forward scattering. 
In contrast,  optical coherence tomography (OCT) leverages reflection-mode measurements and backscattered signals to achieve high sensitivity to lateral interfaces~\cite{huang1991optical}. However,  OCT lacks the ability to quantitatively reconstruct the RI distribution, limiting its utility in applications where detailed RI mapping is critical for understanding sample composition and structure~\cite{drexler2014optical, uttam2015fourier}.

\begin{figure}[t!]
\hspace{-2cm}
\includegraphics[width=17cm]{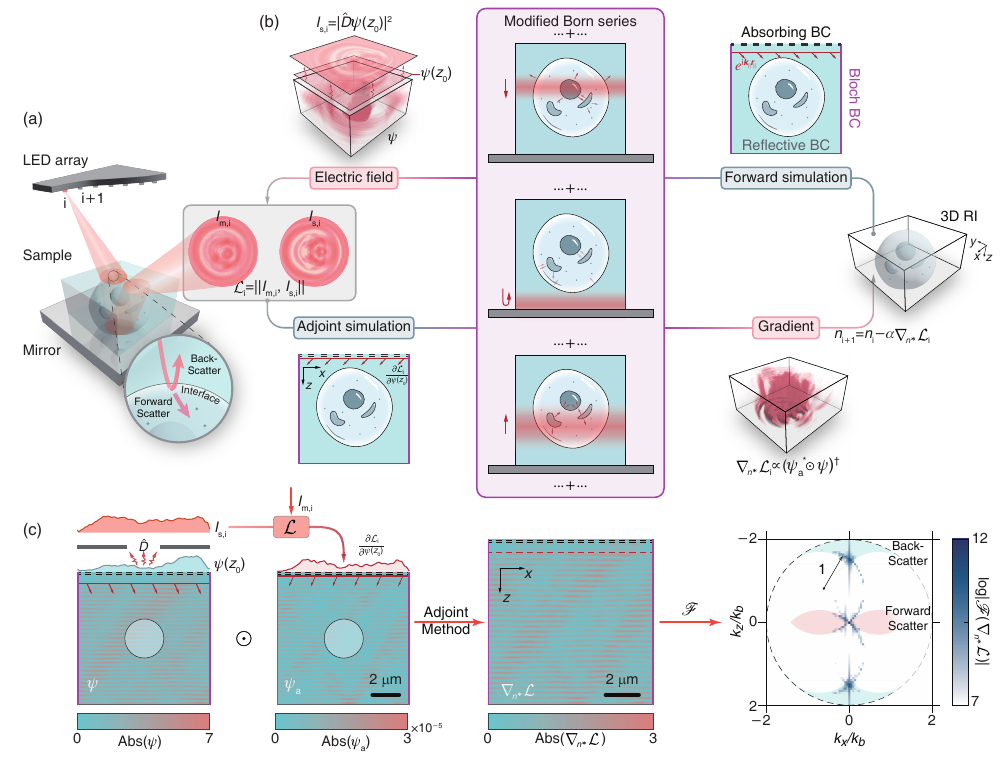}
\caption{\label{Fig1} 
Overview of the reflection-mode DT technique. (a)~A reflection-mode LED array microscope captures brightfield intensity measurements from multiple illumination angles on a sample placed on a reflective substrate. Inset, the diagram of FS and BS in the scattering process. (b)~The reconstruction algorithm iteratively estimates the 3D RI distribution through forward and backward processes. The forward process uses the modified Born series to simulate multiple scattering with appropriate boundary conditions, while the adjoint method computes the gradient in the backward process. (c)~Illustration of AM used for calculating the gradient in a 2D simulation of scattering from a circular scatterer placed on a mirror. Fourier spectrum of the calculated gradient, with orange and blue masks indicating FS and BS supports predicted by the single-scattering model, respectively.
}
\end{figure}

The goal of this work is to bridge this gap by developing a new reflection-mode DT technique. 
Our approach works by imaging a sample on a highly reflective substrate, similar to \cite{unger2019versatile,foucault2019versatile}, combined with a novel reconstruction algorithm, to enable the simultaneous and accurate recovery of both forward scattering (FS) and backward scattering (BS) information (Fig.~\ref{Fig1}(a)).

A key challenge in recovering FS and BS information from reflection-mode DT measurements is the accurate and efficient modeling of complex multiple scattering processes.
Traditional transmission-mode DT techniques typically rely on either single-scattering approximations~\cite{wolf1969three, ling2018high}, which neglect multiple scattering altogether, or multi-slice models~\cite{kamilov2015learning, tian20153d, chen2020multi, zhu2022high}, which can only efficiently model FS. 
To address these limitations, discrete dipole approximation (DDA)~\cite{mudry2010mirror, zhang2013full} discretizes the sample into a finite set of dipoles, enabling the modeling of multiple scattering interactions with improved accuracy. However, its computational cost scales rapidly with the number of dipoles, making it inefficient for imaging large or complex structures with high refractive index contrast. Our previous work on a single-scattering model for reflection-mode DT~\cite{matlock2020inverse} was computationally efficient but limited to thin, weakly scattering samples.

In this work, we overcome these limitations by introducing several key innovations, including the modified Born series (MBS)~\cite{osnabrugge2016convergent} as the forward model, Bloch boundary condition (BC) and perfect electric conductor (PEC) BC to handle oblique incidence and substrate reflections, and the adjoint method (AM)~\cite{lalau2013adjoint,piggott2015inverse, unger2019versatile} for efficient gradient computation in solving the inverse-scattering problem (Fig.~\ref{Fig1}(b)).

Our technique builds on the emerging approach of ``intensity diffraction tomography''~\cite{tian20153d, chen2020multi, zhu2022high}, eliminating the need for a dedicated interferometric setup. Instead of capturing interferograms to retrieve full-field information, our method addresses the inverse-scattering problem using ``phaseless'' intensity measurements collected from multiple illumination angles.

We validate our technique through both simulations and experiments. 
In simulation, we demonstrate the importance of accurately modeling multiple scattering to achieve accurate reconstructions in samples with high RI contrasts and complex geometries. We analyze the complementary roles of FS and BS and develop a straightforward approach for independently accessing their information in the 3D Fourier space. Our results reveal that BS predominantly provides interfacial information, complementing the axially smooth morphological information captured by FS.
In experiment, we quantitatively characterize the imaging performance of our technique using a dual-layer resolution target and 3D randomly distributed beads. We further demonstrate its versatility by imaging a diverse set of 3D samples, including high-resolution phase structures obscured by highly scattering fibers, fixed breast cancer cells, and fixed \emph{C. elegans}.

These results highlight the potential of our technique for high-resolution 3D reconstructions in challenging, strongly scattering environments, with promising applications in semiconductor metrology and inspection, photonic device characterization, and biomedical imaging.

\section{Theory and method}
\subsection{Overview of our reflection-mode DT technique}

Our reflection-mode DT technique aims to bridge the gap between transmission-mode DT and OCT, by simultaneously and quantitatively characterizing the 3D FS and BS information of a sample placed on a highly reflective substrate, as illustrated in Fig.~\ref{Fig1}(a). 
With the assistance of the reflective substrate, such as a silver mirror, the measured scattered field contains both forward- and back-scattered signals, each conveying complementary information about the sample, as shown in the zoomed-in inset of Fig.~\ref{Fig1}(a).
To perform the reconstruction, we first take brightfield intensity images under varying illumination angles \( I_{\text{m},\text{i}} \). 

Once the intensities are collected, the 3D FS and BS information is retrieved by solving the inverse scattering problem to reconstruct the RI distribution, as shown in the iterative loop in Fig.~\ref{Fig1}(b). This process involves two main stages: the forward and backward processes.

In the forward process, MBS serves as an accurate and fast-converging forward model, which has demonstrated its state-of-the-art performance in transmission-mode DT  for complex biological samples~\cite{lee2022inverse}. Starting with an initial guess of the RI (typically the background RI), MBS iteratively refines the electric field \( \boldsymbol{\psi} \), incorporating multiple scattering and handling reflections from the substrate. After convergence, the field on the simulation domain’s surface (\( z = z_0 \)) is projected onto the camera plane through a detection operator \( \hat{D} \). The simulated intensity \( I_{\text{s,i}} \) is compared with the measured data, and the loss function \( \mathcal{L}_\text{i} \) is computed.

In the backward process, AM is used to compute the RI gradient for updating its distribution. An adjoint simulation is performed, generating an adjoint field \( \boldsymbol{\psi}_a \) with a mirrored incident direction. The RI gradient is obtained by element-wise multiplication of \( \boldsymbol{\psi}_a \) and \( \boldsymbol{\psi} \). 
An example of the calculated gradient in a 2D simulation is shown in Fig.~\ref{Fig1}(c). 
Due to the capability of MBS to accurately model the complex multiple scattering processes, the Fourier spectrum of the computed gradient demonstrates that both FS and BS information are efficiently extracted during the reconstruction process, inherently concentrating in the regions around FS and BS supports predicted by the single-scattering model.

After each gradient descent step, the updated RI is used for the next iteration with new measurements under different illumination conditions, continuing until the loss function converges.
Following reconstruction, an additional processing step is applied to the Fourier spectrum of the reconstructed RI to isolate the corresponding regions of FS and BS information for further analysis.
This outlines the main workflow of our reflection-mode DT reconstruction strategy, with further details elaborated in the following sections. 

\subsection{Modified Born series (MBS)}
MBS is an efficient and rigorous forward model for solving the inhomogeneous vectorial Helmholtz equations, which can be expressed as, 
\begin{equation}\label{Dyad_Helm}
    \nabla\times\nabla\times\boldsymbol{\psi}(\boldsymbol{r})-k^2(\boldsymbol{r})\boldsymbol{\psi}(\boldsymbol{r}) =\boldsymbol{S}(\boldsymbol{r}),
\end{equation}
where $\boldsymbol{\psi}(\boldsymbol{r})$ denotes the electric field at spatial position $\boldsymbol{r} \equiv (\boldsymbol{r}_{\parallel}, z)$, $\boldsymbol{S}(\boldsymbol{r})$ is the electric current source at wavelength $\lambda$ and $k(\boldsymbol{r}) = k_0n(\boldsymbol{r})$ is the wavenumber in an isotropic medium with $n(\boldsymbol{r})$ representing the RI distribution, and $k_0=2\pi/\lambda$ is the vacuum wavenumber. By using Green's function, its solution can be formulated as,
\begin{equation}\label{Dya_Helm_solution_mat} \boldsymbol{\psi}(\boldsymbol{r})=\overset{\leftrightarrow}{\mathscr{G}}V(\boldsymbol{r})\boldsymbol{\psi}(\boldsymbol{r})+\overset{\leftrightarrow}{\mathscr{G}}\boldsymbol{S}(\boldsymbol{r}),
\end{equation}
where $\overset{\leftrightarrow}{\mathscr{G}}\equiv\mathscr{F}^{-1}\big[\tilde{\overset{\leftrightarrow}{\boldsymbol{g}}}(\boldsymbol{q})\cdot\mathscr{F}\left(\cdot\right)\big]$ is the Green's function operator, implemented via the angular spectrum method. 
$V(\boldsymbol{r}) \equiv k(\boldsymbol{r})-k_b^2-i\eta$ is the scattering potential, where $k_b\equiv n_bk_0$ is the wavenumber of the background medium with RI $n_b$, and $\eta$ denotes its loss. 
Notably, the choice of $n_b$ and $\eta$ does not affect the solution of Eq.~\eqref{Dyad_Helm}. 
$\mathscr{F}$ and $\mathscr{F}^{-1}$ denote the Fourier and inverse Fourier transforms.
The dyadic Green's function in the background medium has the Fourier transform $\tilde{\overset{\leftrightarrow}{\boldsymbol{g}}}=[\overset{\leftrightarrow}{\boldsymbol{I}}-\boldsymbol{q}\boldsymbol{q}^T/(k_b^2+i\eta)]/(|\boldsymbol{q}|^2-k_b^2-i\eta)$, where $\overset{\leftrightarrow}{\boldsymbol{I}}$ is the identity tensor, and $\boldsymbol{q}$ is the spatial frequency coordinate. 

Expanding $\boldsymbol{\psi}(\boldsymbol{r})$ based on Eq.~\eqref{Dya_Helm_solution_mat} recursively leads to the conventional Born series,
\begin{equation}\label{ConventionalBorn}
    \boldsymbol{\psi}_\mathrm{B}(\boldsymbol{r})=[\overset{\leftrightarrow}{\boldsymbol{I}}+\overset{\leftrightarrow}{\mathscr{G}}V(\boldsymbol{r})+\overset{\leftrightarrow}{\mathscr{G}}V(\boldsymbol{r})\overset{\leftrightarrow}{\mathscr{G}}V(\boldsymbol{r})+\cdots]\overset{\leftrightarrow}{\mathscr{G}}\boldsymbol{S}(\boldsymbol{r}).
\end{equation}
However, the arbitrary choice of $\eta$ in conventional Born series causes serious divergence issues during recursions~\cite{van1999multiple}. 
To address this, G. Osnabrugge et al. proposed a selection strategy for $\eta$, where $\eta$ is chosen as $\eta \ge\max|k^2(\boldsymbol{r})-k_b^2|$ to reduce the influence radius of the Green's function from the original entire simulation domain to a spherically decaying region, ensuring the stability across all scattering regions. 
The effect of $\eta$ is illustrated by the red band matching as the series order increases in Fig.~\ref{Fig1}(b). 
Accordingly, the energy loss is compensated by the $\eta$ term in the scattering potential $V(\boldsymbol{r})$.
A preconditioner $\gamma(\boldsymbol{r})= iV(\boldsymbol{r})/\eta$ is then applied to modify Eq.~\eqref{ConventionalBorn}, yielding the recursive form of MBS:
\begin{equation}\label{ModifiBornsolution}  
\boldsymbol{\psi}_\mathrm{M}(\boldsymbol{r})=[\overset{\leftrightarrow}{\boldsymbol{I}}+\overset{\leftrightarrow}{\mathscr{M}}+\overset{\leftrightarrow}{\mathscr{M}}\overset{\leftrightarrow}{\mathscr{M}}+\cdots]\gamma(\boldsymbol{r})\overset{\leftrightarrow}{\mathscr{G}}\boldsymbol{S}(\boldsymbol{r}),
\end{equation}
where $\overset{\leftrightarrow}{\mathscr{M}}\equiv\gamma(\boldsymbol{r})\overset{\leftrightarrow}{\mathscr{G}}V(\boldsymbol{r})-\gamma(\boldsymbol{r})+\overset{\leftrightarrow}{\boldsymbol{I}}$ (see mathematical details in Section 1 in Supplement 1).

\begin{figure}[ht!]
\hspace{-2cm}
\includegraphics[width=17cm]{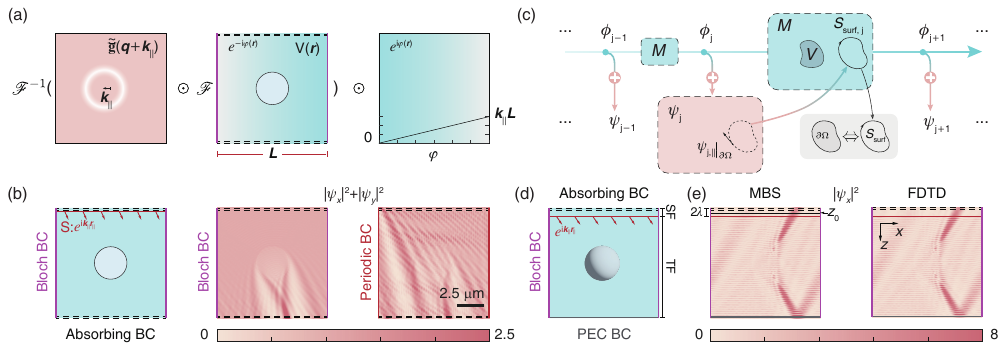}
\caption{\label{Fig2} (a) Padding-free Bloch BC implementation in MBS: 1) apply a phase ramp $e^{-i\boldsymbol{k}_{\parallel}\cdot\boldsymbol{r}_{\parallel}}$ to the entire simulation domain, 2) compute the scattered field using a frequency-shifted Green's function $\tilde{\overset{\leftrightarrow}{\boldsymbol{g}}}(\boldsymbol{q}+\boldsymbol{k}_\mathrm{\parallel})$, 3) re-modulate with a reverse phase ramp $e^{i\boldsymbol{k}_{\parallel}\cdot\boldsymbol{r}_{\parallel}}$ in real space.
(b) Comparison of 2D simulations between Bloch and periodic BCs under oblique incidence. Simulation parameters: $L=10$ $\mu$m, $\lambda = 532$ nm, $n_b = 1.336$; the scatterer has a radius of 1.5 $\mu$m with a RI of 1.461. (c) Computational graph for implementing PEC BC in MBS, realized by a surface current source $\boldsymbol{S}_{\text{surf}}$. (d) MBS simulation domain setup. TF: total field, SF: scattered field. The \( z_0 \) plane is positioned midway between the top absorbing boundary and the one-way source for extracting the scattered field. (e) Comparison of 3D simulations between MBS and FDTD with matching BCs.}
\end{figure}

\subsection{Bloch boundary condition for MBS}
To replicate real experimental conditions within a finite simulation domain, BCs are commonly applied to prevent artificial reflections and unwanted scattering.
A widely used method is the absorbing BC, which absorbs the light waves as they exit the simulation domain in an open system. 
In MBS, absorbing BCs have been implemented using ultra-thin boundary layers~\cite{osnabrugge2021ultra}, offering low computational and memory costs.
To extend MBS for reflection-mode conditions, we further introduce two additional BCs, including Bloch BC (Fig.~\ref{Fig2}(a)) and PEC BC (Fig.~\ref{Fig2}(c)).

Bloch BC~\cite{pendry1996calculating} is commonly used in numerical methods like FDTD to handle the oblique incidence wave $e^{i\boldsymbol{k}_\mathrm{in}\cdot\boldsymbol{r}}$, where $\boldsymbol{k}_\mathrm{in} = [\boldsymbol{k}_{\parallel},k_z]$. 
In FDTD, implementing Bloch BC is straightforward: a constant Bloch phase factor, \( e^{\pm i \boldsymbol{k}_{\parallel} \cdot \boldsymbol{L}} \), is first applied to the boundary field before copying it to the opposite side. Here, \( \boldsymbol{L} \) represents the dimensions of the simulation domain, and the \( + \) and \( - \) signs correspond to the right and left boundaries, respectively. This ensures the correct phase shift for waves exiting and re-entering the simulation domain.

The most straightforward way of applying the Bloch BC based on the angular spectrum method, with a similar way as in FDTD, requires first padding regions at least the same size as the simulation domain before applying the Bloch phase factor to the padded region, in order to both ensure the correct phase shift for waves exiting and re-entering the simulation domain (see details in Section 2 in Supplement 1).
However, this approach results in too much memory overhead. 

To overcome this,  we propose adapting the acyclic convolution (ACC) method~\cite{radhakrishnan2010modified,osnabrugge2021ultra} to achieve a {\it padding-free} Bloch BC in MBS, as illustrated in Fig.~\ref{Fig2}(a). 
First, a phase ramp $e^{-i\boldsymbol{k}_{\parallel}\cdot\boldsymbol{r}_{\parallel}}$ is applied to the entire simulation domain in the real space to effectively demodulate the field before performing discrete Fourier transform (DFT). Correspondingly, we use a frequency-shifted Green's function $\tilde{\overset{\leftrightarrow}{\boldsymbol{g}}}(\boldsymbol{q}+\boldsymbol{k}_\mathrm{\parallel})$ to compute the scattered field.  Finally, the scattered field is re-modulated by $e^{i\boldsymbol{k}_{\parallel}\cdot\boldsymbol{r}_{\parallel}}$ in the real space.
Accordingly, the modified Green's function operator is
\begin{equation}\label{ACC_Bloch}
    \overset{\leftrightarrow}{\mathscr{G}}_\mathrm{Bl} \equiv \mathscr{F}_\mathrm{Bl}^{-1}\big[\tilde{\overset{\leftrightarrow}{\boldsymbol{g}}}(\boldsymbol{q}+\boldsymbol{k}_{\parallel})\cdot\mathscr{F}_\mathrm{Bl}\left(\cdot\right)\big],
\end{equation}
where $\mathscr{F}_\mathrm{Bl}\left(\phi\right)\equiv\mathscr{F}\left(\phi e^{-i\boldsymbol{k}_{\parallel}\cdot\boldsymbol{r}_{\parallel}}\right)$ and $\mathscr{F}_\mathrm{Bl}^{-1}(\tilde{\phi})\equiv\mathscr{F}^{-1}(\tilde{\phi} )e^{i\boldsymbol{k}_{\parallel}\cdot\boldsymbol{r}_{\parallel}}$.
This method ensures that the fields leaking from the boundaries automatically satisfy the Bloch BC, without the need for additional padding or calculations. 
In Fig.~\ref{Fig2}(b), 2D MBS simulations on a circular scatterer with Bloch and periodic BCs are compared. 
The periodic BC introduces strong artifacts due to the mismatch between the incident field at the two ends. 
In contrast, our proposed padding-free Bloch BC ensures that the re-entering field seamlessly matches the internal field.
Note that ACC is not limited to implementing Bloch BC in MBS; it can also be easily extended to other forward models based on the angular spectrum method, such as angular spectrum diffraction, single-scattering, and multi-slice methods.

\subsection{Perfect Electric Conductor (PEC) Boundary Condition for MBS}
We implemented a self-adaptive PEC boundary in MBS to simulate strong reflections from a substrate. While MBS can accurately model reflection at an interface, incorporating a high RI substrate typically requires finer grids and smaller propagation steps, significantly increasing computational complexity. To simplify this, we use a PEC boundary as a lossless reflection model for the substrate interface, providing insights into reflection-mode DT without excessive computational demands.

Based on the equivalence principle~\cite{balanis2012advanced}, a PEC boundary is equivalent to a surface electric current source $\boldsymbol{S}_{\text{surf}}$ on the boundary surface $\partial\Omega$. 
Therefore, applying a PEC BC is converted to solving for $\boldsymbol{S}_{\text{surf}}$ such that $\hat{\boldsymbol{n}}\times\boldsymbol{\psi}(\boldsymbol{r})\big|_{\partial\Omega}=0$.

At each MBS iteration, we incrementally add a surface current source $\boldsymbol{S}'_{\text{surf,j}}\propto \hat{\boldsymbol{n}}\times\boldsymbol{\psi}(\boldsymbol{r})\big|_{\partial\Omega}$, as shown in Fig.~\ref{Fig2}(c). 
The recursive formula for the $j$-th MBS scattering field $\boldsymbol{\phi}_{\text{j}}$ can be written as
\begin{equation}
\boldsymbol{\phi}_{\text{j}+1}=
\overset{\leftrightarrow}{\mathscr{M}}\boldsymbol{\phi}_{\text{j}}+i\boldsymbol{\gamma}\overset{\leftrightarrow}{\mathscr{G}}\left[\boldsymbol{\psi}_{\text{j}}\big|_{\partial\Omega}-\hat{\boldsymbol{n}}\cdot\boldsymbol{\psi}_{\text{j}}\big|_{\partial\Omega}\right],
\end{equation}
where $\hat{\boldsymbol{n}}$ is the normal vector of $\partial\Omega$ and $\boldsymbol{\psi}_{\text{j}}$ is the total field at the $j$-th iteration. 
By defining the added surface current source as $\boldsymbol{S}_{\text{surf,j}}(\boldsymbol{r}) = i\left[\boldsymbol{\psi}_{\text{j}}\big|_{\partial\Omega}-\hat{\boldsymbol{n}}\cdot\boldsymbol{\psi}_{\text{j}}\big|_{\partial\Omega}\right]$, 
the induced electric field cancels the tangential components of the total field estimated on the PEC surface.
Moreover, when the PEC BC is fully satisfied, the added source will vanish, making this a self-adaptive mechanism for integrating the PEC BC in MBS (see mathematical details in Section 2 in Supplement 1). 

\subsection{MBS computational domain setup}

A typical simulation domain in our framework is shown in Fig.~\ref{Fig2}(d). Four side walls are configured with Bloch BCs (purple lines), the bottom plane is assigned a PEC BC (gray line), and the top plane is set with an absorbing BC (double dashed line).

In reflection mode, unlike transmission mode, incident and scattered wave components intermingle within the simulation domain, creating an axial standing wave, as illustrated in Fig.~\ref{Fig2}(d). This overlap complicates field analysis, making it essential to isolate the scattered field from the total field for accurate interpretation. To achieve this, the simulation is divided into distinct regions for the total and scattered fields.

A one-way source is introduced to isolate the scattered field by generating a plane wave propagating in a single direction. This is achieved using a dual-layer source (red line), consisting of two staggered planar sources placed one grid point apart in the \( z \)-direction with an interval \( \Delta z \). In the lateral direction, each planar source has a phase distribution \( e^{i\boldsymbol{k}_{\parallel} \cdot \boldsymbol{r}_{\parallel}} \) to produce the desired oblique incidence. The bottom source also includes an additional phase factor \( -\exp(-ik_z \Delta z) \), canceling any backward-propagating incident waves.
The planar source is positioned two wavelengths below the absorbing boundary, with its \( k_z \) aligned along the positive \( z \)-axis. 

The region between the source and the PEC boundary serves as the total-field region, where structures are placed to simulate scattering. 
The area between the source and the absorbing boundary is the scattered-field region, containing only the pure scattered field from the structure and substrate. 
The $z_0$ plane is positioned in the middle of this gap to capture the scattered field, which is then projected onto the camera plane. The two-wavelength gap effectively eliminates potential evanescent waves near the source or boundary, allowing for accurate analysis of the scattered field.

To validate our method, we simulate the scattering field of a 3D sphere positioned near a PEC boundary, and benchmark the result against FDTD. A spherical scatterer, with a radius of 1.5 $\mu$m and a RI of 1.461, is placed 6 $\mu$m above the PEC boundary within a background medium of RI 1.336. The simulation domain, measuring 12 $\mu$m, is discretized into 20 nm voxels for both MBS and FDTD simulations.
The scatterer is illuminated by a circularly polarized light with a wavelength of 532 nm at an incident angle of \( 30^\circ \). Fig.~\ref{Fig2}(e) compares the \( x \)-\( z \) profiles of the simulated electric fields obtained from MBS and FDTD (additional comparisons are provided in Section 2 in Supplement 1). The simulations show strong agreement, capturing the full scattering lobes, reflections from the substrate, half-wave loss, and standing wave patterns in both cases. 

\subsection{Detection Operator}

After the forward process of MBS, the calculated scattered field at the top slice, \( \boldsymbol{\psi}_M(z_0) \), is projected onto the camera plane using a detection operator \( \hat{D} \), as illustrated in Fig.~\ref{Fig1}(c). The detection operator \( \hat{D} \) models the imaging process in the microscopy system to simulate the captured intensity \( I_{\text{s,i}} \). 
First, \( \boldsymbol{\psi}_M(z_0) \) is numerically propagated to the focal plane of the objective lens and modulated by its pupil function, which ideally acts as a hard aperture with a radius defined by the NA. Finally, the intensities of each polarized component are summed to produce the simulated intensity \( I_{\text{s,i}} \) (see details in Section 3 in Supplement 1).

\subsection{Adjoint method}
To solve the inverse scattering problem, we define a loss function $\mathcal{L}_\text{i}$ that quantifies the difference between the simulated intensity $I_{\text{s,i}}$ and the measured intensity $I_\text{m}$ for each LED illumination:
\begin{equation}
    \mathcal{L}_\text{i} = \left| \left|I_{\text{s,i}} - I_{\text{m,i}}   \right| \right|_2^2,
\end{equation}
where $\parallel\cdot\parallel_2$ denotes the $\ell_2$-norm.
Since MBS is a nonlinear forward model, we minimize 
$\mathcal{L}_\text{i}$ using a gradient descent approach.

In Wirtinger calculus~\cite{wirtinger1927formalen}, the gradient \( \nabla_{n^*} \mathcal{L}_\text{i} \) is computed to update the RI distribution in order to minimize the real-valued loss function \( \mathcal{L}_\text{i} \). To reduce memory costs associated with storing intermediate fields, we employ AM rather than traditional chain-rule-based methods, which has been applied in areas like nanophotonic inverse design~\cite{molesky2018inverse} and DT~\cite{unger2019versatile}. The gradient with AM in the discrete form can be expressed as
\begin{equation}\label{AMM}
    \nabla_{n^*}\mathcal{L}_\text{i}=\left[2k^2_0\text{diag}(\boldsymbol{n})\cdot\boldsymbol{\psi}_\mathrm{a}^*\odot\boldsymbol{\psi}_\mathrm{M}\right]^\dagger,
\end{equation}
where \( \boldsymbol{\psi}_\mathrm{a} \) is the solution to the adjoint simulation, given by
\begin{equation}\label{Adjoin_Helm}
    \nabla\times\nabla\times\boldsymbol{\psi}-k_0^2\text{diag}(\boldsymbol{\epsilon}^*)\boldsymbol{\psi} = \left(\frac{\partial\mathcal{L}_\text{i}}{\partial\boldsymbol{\psi}_\mathrm{M}} \right)^\dagger,
\end{equation}
This adjoint simulation of MBS (see details in Section 4 in Supplement 1) only requires the field at the \( z_0 \)-plane to define \( \mathcal{L}_\text{i} \), making \( (\partial \mathcal{L}_\text{i} / \partial \boldsymbol{\psi}_\mathrm{M})^\dagger \in \mathbb{C}^{N \times 3} \) a plane-shaped source located at the \( z_0 \)-plane. 
If Bloch BCs are used in the forward simulation, the Bloch wave vector must be reversed in the adjoint simulation.

Since AM is derived from the Helmholtz equation (Eq.~\eqref{Dyad_Helm}), Eq.~\eqref{AMM} serves as a universal formula for calculating gradients across all forward models.
Moreover, Eq.~\eqref{AMM} defines the minimal information required to calculate the gradient in a scattering problem. 
While gradients in MBS can be computed using chain-rule-based backpropagation, this approach requires storing all intermediate fields generated during each MBS iteration, resulting in significant memory overhead, especially given MBS's volumetric iterative nature.
In contrast, using AM for gradient calculation reduces memory demands, as only the final scattering field \( \boldsymbol{\psi}_\mathrm{M} \) needs to be retained. This leads to substantial memory savings. 

For validation, we use AM to compute the gradient in a 2D simulation. 
In the simulation, a 2D circular scatterer, with a radius of 1.5 \(\mu\)m and a RI of 1.361, is placed 5 \(\mu\)m above the PEC boundary within a background RI \( n_b = 1.336 \). 
The scatterer is illuminated by a circularly polarized light with a wavelength of 632 nm and an incident angle of \( 40^\circ \).
A pupil with a NA of 0.95 in air is used to capture the scattered field.
The simulation domain has a size of \( L = 10 \) \(\mu\)m, and is discretized into 20 nm pixels to simulate both the forward and backward process.
For comparison, the gradient is also calculated using the chain rule, implemented via PyTorch’s automatic differentiation module (see details in Section 4 in Supplement 1).
Both simulations were run on an Nvidia RTX 4090 24 GB graphics card to compare memory and time consumption.
The gradient results indicate that the gradients calculated using AM closely align with those from the chain rule. However, the difference in computational cost is significant: AM requires only 504 MB of memory and completes in 157 ms, while the chain-rule-based method, which must store all intermediate fields, consumes 20.29 GB and takes 527 ms.

\section{Results}
\subsection{Information Analysis in FS and BS}

Before performing reconstruction tasks, we first analyze the information extracted by reflection-mode DT through interpreting the gradient derived from the loss function. In real space, the gradient reveals lateral stripes with strong oscillations along the axial direction. These axial oscillations correspond to two bright regions in the high axial-frequency domain around \( k_z = \pm 2k_b \) in the Fourier spectrum, as shown in Fig.~\ref{Fig1}(c).

To guide our interpretation, we plot the Fourier support based on the single-scattering model~\cite{jin2017tomographic}, with orange and blue masks representing FS and BS contributions, respectively. 
Although our model includes multiple scattering, which includes higher-order effects, the alignment of the high axial frequency components with the BS region predicted by the single-scattering model indicates that the observed axial oscillations originate from captured backscattered signals. 
Notably, our multiple scattering model also predicts frequency components extending beyond the single-scattering region, as seen in Fig.~\ref{Fig1}(c).
In contrast, FS corresponds to the slowly varying components in the gradient, aligning with the low-frequency region predicted by the single-scattering model, while multiple scattering also introduces additional frequency components extending beyond this region. 

\subsection{Numerical Validation}

We first validate our reflection-mode DT technique in simulation by using gradient descent to reconstruct the 3D RI distribution from simulated intensity measurements. 
In our simulations, we used a light with a wavelength of 632 nm and a pupil with a finite NA of 0.95.
Circular polarization was used in both simulation and experiments to mimic the isotropic and unpolarized illumination from the LED.

\begin{figure}[ht!]
\centering
\includegraphics[width=8.5cm]{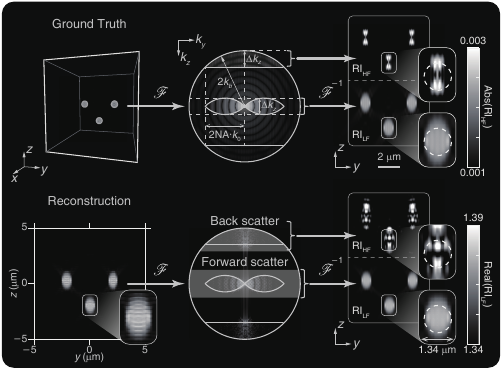}
\caption{\label{Fig3} Reflection-mode DT reconstruction results for simulated beads (RI $n = 1.39$, radius 1 $\mu$m ). Top, ground truth. Bottom, reconstruction results. Simulation parameters: \( \lambda = 632 \) nm, \( n_b = 1.34 \), NA = 0.95. A total of 138 illumination angles are used, arranged in 7 concentric circles with equal spacing in NA, with the outermost circle corresponding to 0.95 NA. Fourier spectrum of the reconstruction shows that $\text{RI}_\text{LF}$, recovered from the FS, and $\text{RI}_\text{HF}$, recovered from the BS, are naturally separated in the 3D Fourier space. The FS and BS supports predicted by the single-scattering model are marked by white curves, with $\Delta k_z = 2k_b(1-\sqrt{1-(\text{NA}/n_b)^2})$. 
By isolating FS and BS supports in the ground truth and the reconstruction, $y$-$z$ profiles of $\text{RI}_\text{LF}$ and $\text{RI}_\text{HF}$ are obtained by taking an inverse Fourier transform.}
\end{figure}

The reflection-mode DT was first tested on beads, with ground truth and reconstruction shown in the top and bottom panels of Fig.~\ref{Fig3}, respectively. 
Since both FS and BS features are present in the gradient (Fig.~\ref{Fig1}(c)), the reconstructed 3D RI contains both slowly varying features and lateral dense stripes in its $y$-$z$ profile.
We perform the Fourier transform on the recovered RI distribution to show the 3D Fourier spectrum of the reconstructed FS and BS components.
In the $k_y$-$k_z$ cross-section of the 3D Fourier spectrum, the FS and BS components are naturally separated along the $k_z$ direction, concentrating in the low- and high-frequency axial regions.
These regions correspond to the FS and BS supports predicted by the single-scattering model, outlined by white curves, with identical lateral and axial bandwidths.
To analyze the information embedded in these two components, they are separately extracted from the Fourier spectrum, followed by an inverse Fourier transform to visualize their real space distributions in the right panel of Fig.~\ref{Fig3}.

We first analyze the reconstructed low-frequency axial component.
The ground truth of the low-frequency RI distribution ($\text{RI}_\text{LF}$) is obtained by isolating the region corresponding to the FS from the 3D Fourier spectrum.
In extracting $\text{RI}_\text{LF}$ from the reconstructed RI, a low-pass filter along the axial dimension is applied in its 3D Fourier spectrum to isolate only the low-frequency components around the FS support. 
This filter slightly extends beyond the single-scattering support to account for extra contributions from multiple scattering (see details in Section 6 in Supplement 1).
The extracted $\text{RI}_\text{LF}$ from the reconstruction shows a smooth distribution, with the dense stripe features removed.
The $y$-$z$ profile of $\text{RI}_\text{LF}$ illustrates similar reconstruction results as the conventional transmission-mode DT and shows good agreement with the ground truth.
Due to the missing cone problem in FS, the boundaries of the reconstructed $\text{RI}_\text{LF}$ are sharp and distinctly resolved in the lateral direction but elongated and blurred in the axial direction.
To evaluate the reconstruction accuracy after incorporating multiple scattering through MBS, the reconstruction is also performed using the single-scattering model (1st Born approximation) for comparison, as illustrated in Section 6 of Supplement 1. The single-scattering model performs poorly in reconstructing the RI, with particularly degraded performance in samples with high RI contrast and complex 3D geometries.

Similar to the $\text{RI}_\text{LF}$, the high-frequency RI distribution ($\text{RI}_\text{HF}$) in the ground truth is obtained by isolating the BS region in its 3D Fourier spectrum. 
The $\text{RI}_\text{HF}$ in the reconstruction is extracted by applying a high-pass filter at the BS support to isolate the recovered high-frequency components.
High-frequency oscillations with an approximate frequency of $2n_b/\lambda$ along the $z$-direction appear in both the real and imaginary parts of $\text{RI}_\text{HF}$ due to its shift from the zero frequency.
Since our method provides the complex-valued reconstruction of $\text{RI}_\text{HF}$, the envelope of $\text{RI}_\text{HF}$ along $z$ can be obtained by calculating its absolute value, as shown in Fig.~\ref{Fig3}.
In both the ground truth and reconstruction of $\text{RI}_\text{HF}$, strong features are present near the top and bottom surfaces of the beads, while they diminish inside the beads (see details in Section 6 in Supplement 1).
This result suggests that the recovered information from BS is more sensitive to the lateral interfaces (with a normal vector along the $z$-direction) of the scatterers.

\subsection{Experiments}

\begin{figure}[ht!]
\hspace{-2cm}
\includegraphics[width=17cm]{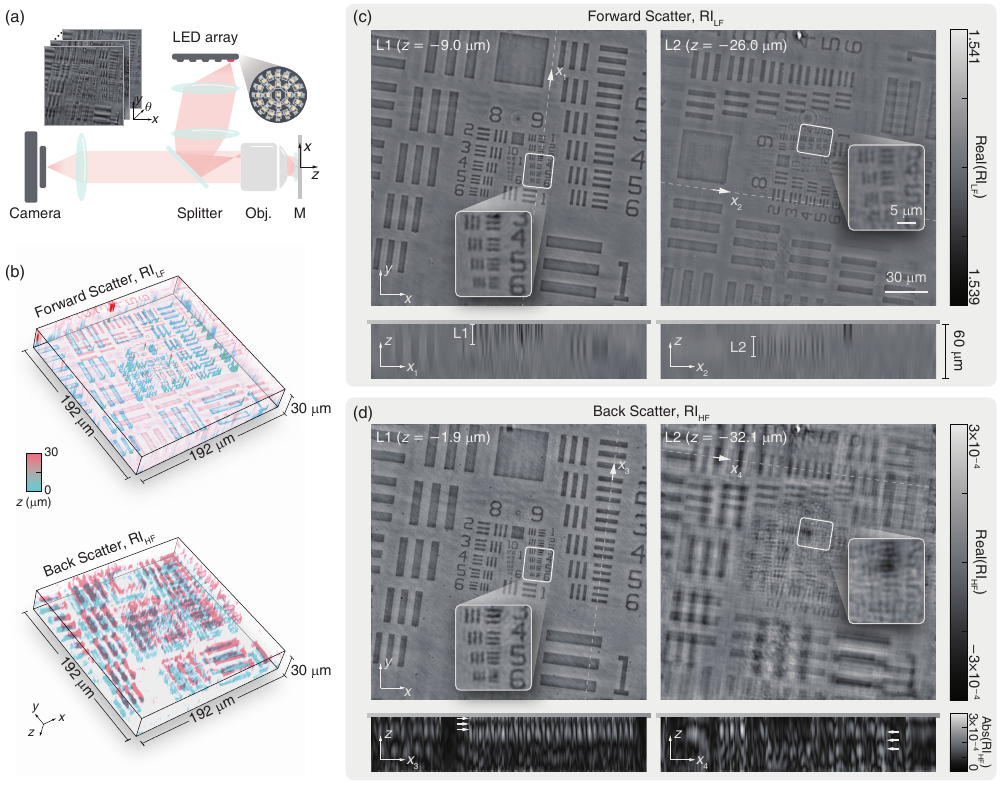}
\caption{\label{Fig4} Imaging of a two-layer resolution target sample.
(a)~Schematic of our reflection-mode DT setup. Obj: objective lens, M: mirror. Illumination is provided by a 25-LED array with a central wavelength of 632 nm. The illumination NAs of the two LED rings are 0.14 and 0.28. Inset: captured 2D intensity images from different illumination angles. 
(b)~3D rendering of reconstructed $\text{RI}_\text{LF}$ and $\text{RI}_\text{HF}$ of the dual-layer resolution target. 
(c)~\(x\)-\(y\), \(x_1\)-\(z\), and \(x_2\)-\(z\) cross-sections of the reconstructed $\text{RI}_\text{LF}$. 
The gray dashed lines, \(x_1\) and \(x_2\), intersect Group 9 in Layer 1 and Group 8 in Layer 2, illustrating the \(z\) cross-sections. 
The zoomed-in inset of the \(x\)-\(y\) cross-sections highlights the reconstructed regions around Element 5 in Group 9 for both layers, demonstrating diffraction-limited lateral resolution in both layers. 
(d)~\(x\)-\(y\), \(x_3\)-\(z\), and \(x_4\)-\(z\) cross-sections of the reconstructed $\text{RI}_\text{HF}$. 
The gray dashed lines, \(x_3\) and \(x_4\), intersect Group 7 in Layer 1 and Layer 2, illustrating the envelopes of $\text{RI}_\text{HF}$ in \(z\) direction.
The zoomed-in inset of the \(x\)-\(y\) cross-section in Layer 1 demonstrates the achievement of diffraction-limited lateral resolution in BS.
The \(x\)-\(y\) cross-section of Layer 2 overlaps with the ghost images from Layer 1.
}
\end{figure}

Our experimental setup is based on our previously developed reflection-mode LED microscope \cite{wang2023fourier}, as shown in Fig.~\ref{Fig4}(a). A 10$\times$/0.28 NA (Mitutoyo Plan Apo Infinity Corrected Long WD) objective lens collects the scattered signals. For oblique illumination, a 25-LED array (Kingbright) is positioned at the focal plane of a 4-$f$ system and relayed to the back focal plane of the objective. The LED array consists of a central LED and two concentric rings, with the outermost ring matching the objective NA, providing illumination NAs of 0.14 and 0.28 for the rings. Each LED illuminates the sample as a plane wave with a central wavelength of 632 nm and a 20 nm bandwidth. The intensity is captured by a camera (Imaging Source, DMK38UX541, 2.74 $\mu$m pixel size). The illumination angle is calibrated using the method from~\cite{eckert2016algorithmic, wang2023fourier}.

\subsubsection{Two-layer phase resolution target}

We first image a dual-layer structure to validate lateral resolution and demonstrate the capability of reflection-mode DT in resolving complex 3D structures. Conventional microscopy struggles with scattering from out-of-focus regions, which obscures structures at different depths. This issue is particularly severe in reflection-mode, where light undergoes both FS and BS events, compounding scattering effects. Our technique overcomes these limitations by reconstructing the 3D structure through solving an inverse scattering problem.

For validation, we fabricated a dual-layer phase resolution target. 
Two distinct targets were replicated in clear resin ($n_b\approx1.54$), with a nominal axial separation of 17.5 $\mu$m. Layer 1 has a nominal height of 200 nm, while Layer 2 is 100 nm.
The master target used for replication is the Quantitative Phase Target (Benchmark Technologies).
The layers were stacked on a silver mirror, aligned at the center. The top layer was rotated 90 degrees, and the space was filled with glycerin ($n\approx1.48$) before sealing with a cover glass.

The raw measurements, shown in the inset of Fig.~\ref{Fig4}(a), include contributions from the field passing through the sample twice, as the substrate reflects it back through the sample. 
This causes significant overlap of phantom images from each layer in the measurement.

The reconstructed volume extends from the mirror surface to 60 $\mu$m above, with a field of view (FOV) of 192 $\mu$m $\times$ 192 $\mu$m. 
It is discretized on a uniform grid with 103 nm spacing in all three dimensions. The 3D rendered RI reconstruction ($z\in[0,30\,\mu\text{m}]$) is shown in Fig.~\ref{Fig4}(b). 
After reconstruction, the $\text{RI}_\text{LF}$ (FS) and  $\text{RI}_\text{HF}$ (BS) volumes are extracted from the Fourier domain, with their cross-sections shown in Fig.~\ref{Fig4}(c) and (d), respectively.

The zoomed-in \(x\)-\(y\) profiles inset in $\text{RI}_\text{LF}$ and $\text{RI}_\text{HF}$ demonstrate a significant enhancement in lateral resolution for both FS and BS reconstructions.
In $\text{RI}_\text{LF}$ (Layer 1 and 2) and $\text{RI}_\text{HF}$ (Layer 1), features in Group 9, Element 5 (1230 nm period) are clearly resolved, as shown in the zoomed-in insets of Fig.~\ref{Fig4}(c) and (d). This confirms that the reconstruction achieves the expected synthetic NA of 0.56 in both FS and BS, corresponding to a theoretical diffraction-limited resolution of 1128 nm, representing a 2$\times$ improvement in NA over single-LED brightfield measurements.

In \(x\)-\(y\) profiles, four gray dashed lines are plotted, intersecting the target elements to illustrate their $z$ cross-sections in the bottom panels.
In $\text{RI}_\text{LF}$, the \(x_1\)-\(z\) and \(x_2\)-\(z\) cross-sections demonstrate the achieved axial resolution in resolving the two-layer structure.
The current axial resolution and RI quantification are limited by the system's NA, similar to the observations in transmission-mode DT \cite{chen2020multi}.
This causes the reconstructed RI distributions of both layers to extend along the $z$-axis, spanning approximately $24.0$ $\mu$m, which aligns with the expected axial elongation from low-frequency band Fourier coverage ($24.6$ $\mu$m).
Since the nominal height of Layer 2 is half that of Layer 1, the $\text{RI}_\text{LF}$ distribution in Layer 2 exhibits lower contrast in the \(x\)-\(y\) profile.
Averaging the $\text{RI}_\text{LF}$ difference between target and background, $\langle\Delta n\rangle=\overline{n-n_b}$, for Group 8 elements, the contrast ratio between Layer 1 and Layer 2,  $\langle\Delta n_1\rangle/\langle\Delta n_2\rangle$, is 1.82, close to the nominal height ratio of 2.

In $\text{RI}_\text{HF}$, the \(x_3\)-\(z\) and \(x_4\)-\(z\) cross-sections, shown in the bottom panel of Fig.~\ref{Fig4}(d), illustrate the $z$-directional envelopes of the reconstructed $\text{RI}_\text{HF}$. 
Gibbs phenomena arise in $z$-directional envelopes due to several factors, including the absence of low-frequency support, the limited axial bandwidth of the BS components, and the sharp $z$-profiles of the targets.
As a result, prominent ghost images from the Layer 1 target appear in the $x$-$y$ profiles of the Layer 2 target, and both targets exhibit shifts in the reconstructed $z$-positions.
Future improvements in enhancing axial resolution of FS and BS, more accurate quantification of $\text{RI}_\text{LF}$, and suppression of ghost images in $\text{RI}_\text{HF}$ can be achieved by increasing the axial bandwidth, e.g., with a higher-NA system.

\subsubsection{3D randomly distributed beads}

\begin{figure}[ht!]
\centering
\includegraphics[width=8.5cm]{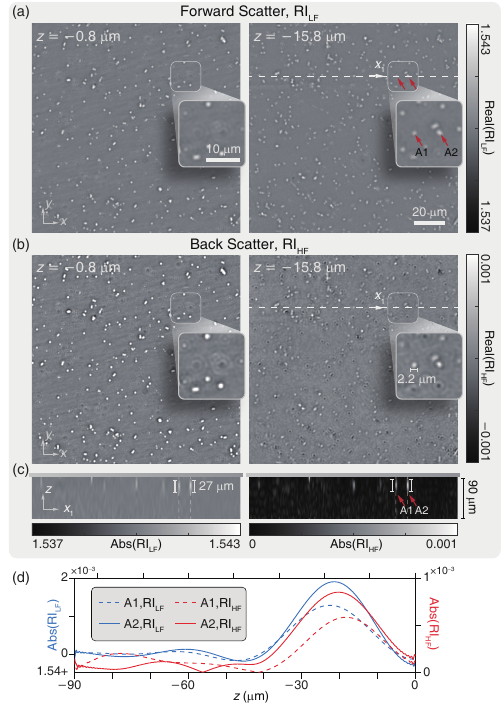}
\caption{\label{Fig5} Imaging of a 3D randomly distributed beads sample. 
(a)~$x$-$y$ cross-sections of $\text{RI}_\text{LF}$ reconstructed from forward-scattered signals at $z=-0.8$ $\mu$m and $z=-15.8$ $\mu$m planes.
The zoomed-in inset shows $x$-$y$ cross-sections of beads A1 and A2, which have approximated diameters of $2.2$ $\mu$m.
(b)~$x$-$y$ cross-sections of $\text{RI}_\text{HF}$ reconstructed from back-scattered signals at $z=-0.8$ $\mu$m and $z=-15.8$ $\mu$m planes.
(c)~The gray dashed line  $x_1$ in (a) and (b) intersect $\text{RI}_\text{LF}$ and $\text{RI}_\text{HF}$ of beads A1 and A2 to show their $z$-directional cross-sections. 
(d)~The gray dashed lines placed at the center of each beads in (c) intersect the $z$ cross-sections of beads A1 and A2 to visualize their $z$-directional envelopes.}
\end{figure}

Next, we image a 3D sample of randomly distributed beads, where the beads are of moderate sizes and exhibit stronger BS signals compared to phase resolution targets.
This makes them ideal for validating the axial resolution of FS and BS in our technique.
We fabricated the sample by dispersing polystyrene spheres ($n\approx1.59$) with an average diameter of $2$ $\mu$m in resin ($n_b\approx1.54$) and fixing the mixture onto the surface of a silver mirror.
The same experimental setup was used to capture intensities at different illumination angles for reflection-mode DT reconstruction.
The reconstructed volume is discretized into 103 nm voxels in all directions and extends from the mirror surface to 90 $\mu$m, with a FOV of 137 $\mu$m $\times$ 137 $\mu$m. 
After reconstruction, $\text{RI}_\text{LF}$ and $\text{RI}_\text{HF}$ are extracted from the Fourier domain for analysis and illustration.

The $x$-$y$ cross-sections of the reconstructed $\text{RI}_\text{LF}$ and $\text{RI}_\text{HF}$ at $z=-0.8$ $\mu$m and $z=-15.8$ $\mu$m are illustrated in Fig.~\ref{Fig5}(a) and (b), respectively.
Both cross-sections reveal the spatial distribution of beads at different depths.
In $x$-$y$ cross-sections of real($\text{RI}_\text{LF}$), beads at different depths appear as white blobs with varying values. In contrast, real($\text{RI}_\text{HF}$) shows beads with not only varying values but also sign differences, with some beads appearing as positive (white) and others as negative (black) due to the oscillatory nature of the high-frequency components.

Beads A1 and A2, with approximated diameters of $2.2$ $\mu$m, as shown in the zoomed-in inset, are randomly selected to quantify the axial resolution.
The gray dashed lines in Fig.~\ref{Fig5}(a) and (b) intersect $\text{RI}_\text{LF}$ and $\text{RI}_\text{HF}$ of beads A1 and A2 to show their $z$ cross-sections in Fig.~\ref{Fig5}(c).
Due to the identical axial bandwidth of the FS and BS supports, $\text{RI}_\text{LF}$ and $\text{RI}_\text{HF}$ exhibit similar axial elongations in the $z$-direction. 
They span approximately $27.0$ $\mu$m in the $z$-direction, close to the expected axial resolution of $24.6$ $\mu$m.
In Fig.~\ref{Fig5}(c), two gray dashed lines are placed at the center of each beads, intersecting the $z$ cross-sections of both beads to visualize their $z$-directional envelopes, shown in Fig.~\ref{Fig5}(d). 
The envelopes of $\text{RI}_\text{LF}$ and $\text{RI}_\text{HF}$ are marked as blue and red curves, respectively.
The limited NA of our systems cause both $\text{RI}_\text{LF}$ and $\text{RI}_\text{HF}$ to extend along $z$-direction, with their envelopes forming bell-shaped curves (see details in Section 7 in Supplement 1). 
These results demonstrate that by analyzing the axial envelope of $\text{RI}_\text{HF}$, FS and BS offer the same axial resolution in reflection-mode DT, with the measured axial resolution agreeing with the theoretical prediction.

\subsubsection{Phase spoke target behind random fibers}
\begin{figure}[ht!]
\centering
\includegraphics[width=8.5cm]{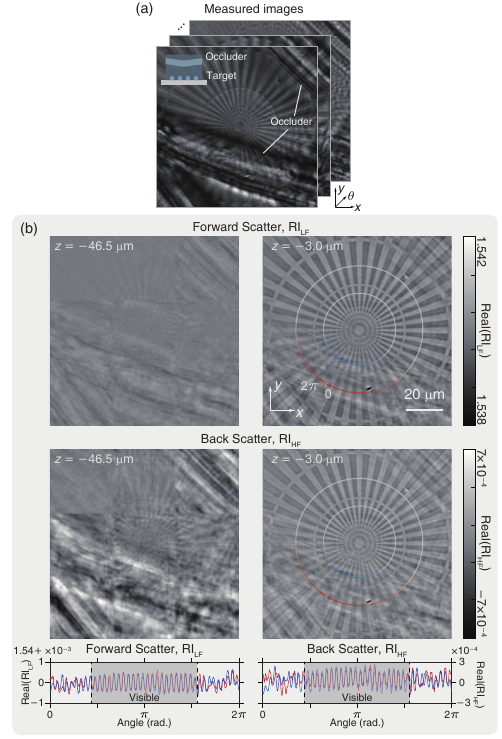}
\caption{\label{Fig6} Imaging of a phase spoke target behind random fibers. (a)~Example measurements under different illumination angles. (b)~$x$-$y$ cross-sections of the reconstructed $\text{RI}_\text{LF}$ and $\text{RI}_\text{HF}$ at $z=-46.5$ $\mu$m and $z=-3.0$ $\mu$m planes. After reconstruction, most of the previously occluded spoke patterns are now clearly resolved in both $\text{RI}_\text{LF}$ and $\text{RI}_\text{HF}$. 
The intense back-scattered signals from the fiber interfaces are reflected in reconstructed $x$-$y$ cross-section of $\text{RI}_\text{HF}$ at $z=-46.5$ $\mu$m, where high-contrast features appear along the edges and tops of the fibers.
Bottom panel, the reconstructed $\text{RI}_\text{LF}$ and $\text{RI}_\text{HF}$ along two circular traces marked in the $x$-$y$ profile. The area between the dashed lines represents the visible region in measurements, while the areas on either side correspond to the occluded regions.}
\end{figure}

After characterizing the performance of our system, we demonstrate its versatility by imaging structures obscured by strongly scattering occluders, simulating challenging metrology tasks in highly scattering environments.
These occluders, such as scratches, dust, or overlapping elements, act as irregular strong scatterers, generating intense back-scattered signals and significantly distorting forward-scattered signals, making direct inspection challenging in some metrology tasks. Reflection-mode DT provides a novel approach for visualizing structures behind these occluders.

To mimic an obscured sample, we use a torn piece of lens tissue as an occluder. Made of coarse fibers, the tissue strongly scatters light, creating a semi-transparent but highly distorted view, making it ideal for this purpose. For sample preparation, a resin-copied spoke resolution target is affixed to a mirror surface, with a piece of lens tissue secured above it using resin. 
Nominal height of the target structure is 300 nm, and the separation between the fiber plane and the target is approximately 45 $\mu$m.

The same experimental setup was used to capture intensities at different illumination angles, with example images shown in Fig.~\ref{Fig6}(a). In the captured data, parts of the target are obscured by random fibers, causing significant scattering and distortion, making the resolution features nearly unobservable.

Reflection-mode DT reconstruction enables separation of the target and occluders along the $z$-direction.
The reconstructed volume extends from the mirror surface to 55 $\mu$m above, with an FOV of 104 $\mu$m $\times$ 104 $\mu$m.
$x$-$y$ cross-sections of the reconstructed $\text{RI}_\text{LF}$ and $\text{RI}_\text{HF}$ at $z=-46.5$ $\mu$m and $z=-3.0$ $\mu$m planes are shown in Fig.~\ref{Fig6}(b). 
At the plane farther from the mirror ($z=-46.5$ $\mu$m), the $x$-$y$ cross-section of $\text{RI}_\text{HF}$ reveals high-contrast features along the edges and tops of the fibers, corresponding to strong BS signals from the fiber surfaces.
At the plane closer to the mirror ($z=-3.0$ $\mu$m), the occluded spoke resolution features are now clearly visible in both $\text{RI}_\text{LF}$ and $\text{RI}_\text{HF}$, with edges aligned seamlessly with uncovered regions.
The reconstructed $\text{RI}_\text{LF}$ and $\text{RI}_\text{HF}$ along two circular traces are plotted at the bottom for detailed comparison, with previously obscured regions highlighted in blue and red in the $x$-$y$ cross-sectional images. 
While not as sharp as uncovered areas, the previously distorted structures are now distinctly visible.

In this reconstruction, 3D total variation (TV) regularization \cite{kamilov2016optical} is applied to suppress reconstruction artifacts and enhance the result. TV regularization is applied only to $\text{RI}_\text{LF}$ distribution after each step of the gradient descent to prevent the influence of $\text{RI}_\text{HF}$. (see details in Section 5 of Supplement 1).

This result underscores the potential of our technique for achieving accurate inspection of structures on substrates obscured by defects or overlapping elements, presenting a promising solution for challenging inspection scenarios.

\subsubsection{Imaging of biological samples}

\begin{figure}[ht!]
\hspace{-2cm}
\includegraphics[width=17cm]{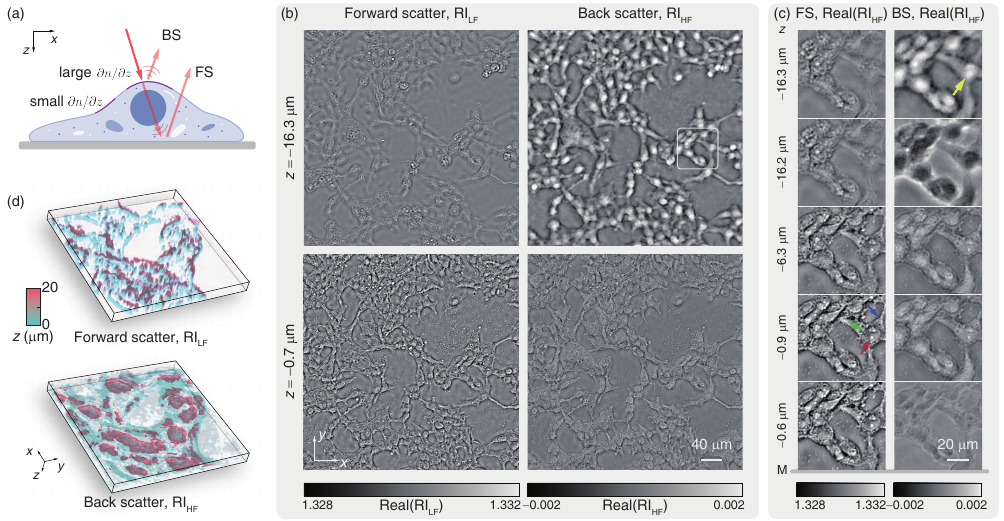}
\caption{\label{Fig7} Imaging of  breast cancer cells fixed on a mirror. (a) Illustrations of the FS and BS in the biological samples. (b) $x$-$y$ cross-sections of $\text{RI}_\text{LF}$ and $\text{RI}_\text{HF}$ at $z=-16.3$ $\mu$m and $z=-0.7$ $\mu$m. (c) Zoomed-in $z$-stack of $x$-$y$ cross-sections of $\text{RI}_\text{LF}$ and $\text{RI}_\text{HF}$ at different $z$ plane. The zoomed-in region is marked with a white frame in (b). (d) 3D rendering of the zoomed-in region of $\text{RI}_\text{HF}$ and $\text{RI}_\text{LF}$. }
\end{figure}

Finally, we apply our technique to image biological samples.
Compared to metrology samples, biological samples have more complex internal structures and varied surface topographies, making them particularly suitable for demonstrating the differences between $\text{RI}_\text{HF}$ and $\text{RI}_\text{LF}$.
When the incidence impinges upon the biological samples, as shown in Fig.~\ref{Fig7}(a), a portion of the wave is directly reflected by the interfaces, contributing to the BS signals, while the remaining part passes through the sample volume twice and is reflected by the substrate, contributing to the FS signals.
By reconstructing $\text{RI}_\text{HF}$ and $\text{RI}_\text{LF}$ from these signals, we can extract complementary information about the sample's structural features.

First, we present results of breast cancer cells fixed on the mirror. Additional results from a fixed \emph{C. elegans} are provided in Section 7 in Supplement 1.
The unstained breast cancer cells were first cultured on the mirror surface, then fixed with formalin and immersed in water during measurement.
The same experimental setup was used.
The reconstructed volume is discretized into 118 nm voxels in all directions, and extends from the mirror surface to 20 $\mu$m above, with an FOV of 411 $\mu$m $\times$ 411 $\mu$m.
After reconstruction, $\text{RI}_\text{LF}$ and $\text{RI}_\text{HF}$ are extracted, with their $x$-$y$ cross-sections at $z=-16.3$ $\mu$m and $z=-0.7$ $\mu$m illustrated in Fig.~\ref{Fig7}(b).
Due to the limited NA of our setup, the $z$-values indicate approximate depths and do not represent the exact thickness of the cells.
A $z$-stack and 3D rendering of the zoomed-in cell clusters are illustrated in Fig.~\ref{Fig7}(c) and (d). 

At the $z=-16.3$ $\mu$m plane, which is away from the mirror surface and approximately intersects the peak parts of the breast cancer cells, the $x$-$y$ cross-sections of $\text{RI}_\text{LF}$ in Fig.~\ref{Fig7}(b) and (c) exhibit blurring and low-contrast cell features.
This is attributed to the missing cone problem in FS, which results in limited axial resolution and low sensitivity to the top water-membranes interfaces. 
In contrast, new features emerge in the reconstructed $\text{RI}_\text{HF}$ that are not visible in $\text{RI}_\text{LF}$, revealing complementary interfacial details of the breast cancer cells.
These high-contrast features, reconstructed from signals back-scattered from the water-membrane interfaces, exhibit strong contrast against the background and demonstrate continuous lateral variations.
Morphologically, the membranes of the fixed breast cancer cells are inherently raised at the nucleus region, forming a smooth interface at the top, as shown in Fig.~\ref{Fig7}(a).
Since the lateral interfaces have steeper $z$-directional gradients compared to the oblique interfaces, the BS signals become more sensitive to these regions. As a result, the reconstructed interfacial features effectively highlight these nuclear regions, which are marked by a yellow arrow in Fig.~\ref{Fig7}(c).

As the intersection approaches the mirror surface at the $z=-0.9$ $\mu$m plane, the $x$-$y$ cross-sections of $\text{RI}_\text{LF}$, illustrated at the bottom of Fig.~\ref{Fig7}(b) and (c), reveal features similar to those seen in transmission-mode DT, where subcellular structures, such as cell edges (blue arrow), nuclear envelope (green arrow) and, internal nucleoli  (red arrow), become clearly visible.
In BS recovery, the high-contrast interfacial features gradually disappear in the $x$-$y$ cross-sections of $\text{RI}_\text{HF}$, while the subcellular structures corresponding to $\text{RI}_\text{LF}$ become more visible.
Since the axial RI contrast between subcellular structures and cytoplasm is smaller than that at the water-membrane interfaces, the contrast of the reconstructed $\text{RI}_\text{HF}$ is lower for subcellular structures compared to the water-membrane interface.

These results demonstrate that the information recovered from both FS and BS in reflection-mode DT can reveal subcellular features, with FS capturing axially smooth variations in cellular structures, and BS providing complementary information about cellular interfaces.
This capability underscores the potential of reflection-mode DT for the simultaneous and independent characterization of complex internal structures and interfacial features of biological samples, offering a powerful approach for detailed biological analysis.

\section{Conclusion and Discussion}
In conclusion, we presented a novel reflection-mode DT technique that utilizes a highly reflective substrate and a novel reconstruction algorithm, enabling the simultaneous reconstruction of both FS and BS information from intensity-only measurements.
By integrating the MBS as the forward model, incorporating Bloch and PEC BCs to handle oblique incidence and substrate reflections, and employing the adjoint method for efficient gradient computation, our proposed algorithm effectively separates and extracts both volumetric and interfacial information from the measurements.
Experimental validation on a reflection-mode LED array microscope demonstrated the capability of achieving high-resolution 3D reconstructions across a diverse range of scattering samples.

These innovations open new avenues for expanding the applications of DT in both metrology and biomedical fields.
Reflection-mode DT is particularly well-suited for scenarios involving strongly scattering samples or those naturally positioned on reflective substrates, such as those encountered in metrology and inspection applications \cite{kim2016large, kang2023accelerated, aidukas2024high}. In industrial contexts, this technique can be applied for high-resolution inspection of semiconductor wafers, nano-structures, and photonic devices, where both FS and BS signals offer complementary insights vital for precise 3D characterization.
In the biomedical domain, reflection-mode DT facilitates new applications in label-free microscopy. For instance, in mid-infrared photothermal microscopy \cite{bai2021bond,zhang2019bond,tamamitsu2020label,zhao2022bond,xia2022mid}, where reflection geometry is preferred, reflection-mode DT allows for the simultaneous quantification of refractive index changes induced by thermal effects in biological samples from the mid-infrared pump beam, while maintaining high sensitivity to surface and interface details.

Future work could focus on improving the resolution and computational efficiency of reflection-mode DT. For FS, axial resolution can only be enhanced by increasing the system’s NA or using shorter wavelengths, as the FS support of a longer wavelength is fully contained within that of a shorter one. In contrast, for BS, axial resolution can also be improved through multi-wavelength illumination, as the BS supports of different wavelengths spread axially in the Fourier domain~\cite{zhang2018multi}.
Moreover, computational costs can be further reduced by developing simplified models optimized for specific scenarios, using the MBS as a benchmark to maintain accuracy. To better integrate FS and BS information, incorporating advanced computational methods, such as deep learning \cite{barbastathis2019use}, could provide an efficient and low-cost approach for achieving 3D isotropic imaging.
These advancements will broaden the applications of reflection-mode DT in both industrial and biomedical fields, offering more advanced, high-resolution, and non-invasive imaging solutions.

\begin{backmatter}
\bmsection{Funding}
Samsung Global Research Outreach (GRO) program, and
National Science Foundation (1846784).

\bmsection{Acknowledgment}
The authors thank Boston University Shared Computing Cluster for proving the computational resources. The authors thank Hongjian He, Dr. Jianpeng Ao, Dr. Dashan Dong and Dr. Ji-xin Cheng for providing the breast cancer cell samples and the \emph{C. elegans} samples. The authors also thank BU CISL group members for insightful discussions.

\bmsection{Disclosures}
The authors declare no conflicts of interest.


\bmsection{Supplemental document}
See Supplement 1 for supporting content.

\end{backmatter}


\end{document}